\documentclass[12pt,a4paper]{article}
\usepackage{amssymb}
\usepackage{amsthm}
\usepackage{amsmath}
\usepackage{epsfig}
\usepackage[latin1]{inputenc}
\usepackage{fancybox}
\usepackage{xspace}

\usepackage{breqn}

\title{Bootstrap testing for cross-correlation under low firing activity}
\date{University of A Coru\~na \\ \vspace{1cm}March, 2015}

\author{Aldana M. Gonz\'alez-Montoro \and
Ricardo Cao \and Nelson Espinosa \and Javier Cudeiro \and
Jorge Mari\~no}

\begin{document}


\maketitle

\begin{abstract}
A new cross-correlation synchrony index for neural activity is proposed. The index is based on the integration of the kernel estimation of the cross-correlation function. It is used to test for the dynamic synchronization levels of spontaneous neural activity under two induced brain states: sleep-like and awake-like. Two bootstrap resampling plans are proposed to approximate the distribution of the test statistics. The results of the first bootstrap method indicate that it is useful to discern significant differences in the synchronization dynamics of brain states characterized by a neural activity with low firing rate. The second bootstrap method is useful to unveil subtle differences in the synchronization levels of the awake-like state, depending on the activation pathway.\\

\end{abstract}

\section{Introduction}
\label{intro}

During deep sleep, neurons are highly synchronized, embedded in slow oscillations (hence the name `slow-wave sleep'). In this period, most neurons of the cerebral cortex display an oscillatory behavior, generating bursts of spikes with a dominant rhythm of about 1-5\,Hz (1--0.20\,s between bursts): the delta rhythm. This oscillation is highly synchronized among neurons in the cortex and other brain regions. Because of this massive synchronization, the global electrical activity displays a high amplitude oscillation that can be easily observed in the electroencephalogram (EEG) as it shows larger amplitude and lower frequency waves than in the awake state (Steriade, McCormick and Sejnowski, 1993). Under experimental conditions this global oscillatory activity can be induced by some anesthetics, giving rise to a sleep-like activity. This allows the study of the neuronal properties of the oscillatory phenomena from an electrophysiological point of view. During the awake state such global oscillatory synchronized activity does not exist, and neuronal spikes are not organized in repetitive bursts of activity, but follow what could be seen as a more random response, generating trains of spikes with different patterns and frequencies . This mode of operation is referred to as tonic activity, in contrast to the mentioned slow oscillatory activity (of course, this tonic activity is not random, and is used to convey all kinds of information). \\

The transition between the sleep and awake states is modulated by the so called `activating ascending pathways' which originate in neuronal nuclei located in the brain stem (\textit{bs}) and basal forebrain (\textit{bf}). Each activating pathway uses specific neurotransmitters, delivered through specific neural routes (Steriade, 1994; Bazhenov et al.,  2002). The experimental electrical stimulation of these nuclei can change the EEG pattern from the typical sleep-like pattern to the one expected in an awake individual. This happens because stimulation suppresses the slow oscillatory activity, and promotes a tonic mode of activity, thus introducing a tool to study the effects of the mechanisms that underlie the sleep-wake cycle (Hu et al., 1989; Burlet et al., 2002; Mari\~no and Cudeiro, 2003). To study this process under controlled laboratory conditions, animals are anesthetized and the activating nuclei manipulated by means of electrical stimulation. Some anesthetics induce a sleep-like state virtually identical to the natural one; under this condition, the electrical stimulation of either the \textit{bs} or the \textit{bf} promotes a change in the global brain activity and, for a period of some seconds after the stimulation (usually from 2 to 20\,s), the EEG shows an awake-like pattern. Thus, it is possible to study the spontaneous synchronization dynamics during those states for the same pairs of neurons, and to look for subtle differences in the awake-like state induced by \textit{bs} and \textit{bf}, but this requires a synchrony measure sensitive enough to deal with low spike activity, together with powerful statistical tests.\\

During sleep-like oscillatory periods (either natural or induced by anesthesia) the activity is dominated by the 1--5\,Hz delta rhythm, in which the firing rate is mostly reduced to brief bursts of activity. There exists also an increase in alpha activity (8--12\,Hz) (Brown et al., 2010). Spontaneous activity is usually characterized by low firing activity. Under this condition, statistical analysis becomes difficult and therefore, in order to make inferences, special measures and appropriate statistical testing procedures are needed. The aforementioned experimental conditions permit the study of dynamic synchronization between pairs of neurons under two types of spontaneous activity: the anesthetic-induced sleep-like activity, and the electrically induced awake-like activity. The awake-like type of activity needs an initial electrical stimulus to be induced. Afterwards, such mode of activity can last for tens of seconds. Hence,  we will consider both the sleep-like and the awake-like signals as periods of spontaneous activity. In any case, this is only a terminological issue that does not change at all the mathematical and statistical results and conclusions.\\

Cross-corre\-la\-tion analysis of simultaneously isolated single-neuron activities is a common procedure to study the degree of synchronization evoked under certain conditions, like sensory stimulation. Those approaches are essential to study the information coding and functional organization of the brain, but the spontaneous spike activity can also provide important clues to brain structure and function.\\

Correlations between spike trains of neuron pairs can modulate neuronal activity and affect how neurons encode information. Furthermore, sensory and cognitive processing relies on the concerted activity of large populations of neurons. For instance, in the visual system  more information can be extracted from the activity of pairs of cells in the lateral geniculate nucleus if correlations between their spikes are taken into account (Dan et al., 1998). However correlated neural activity and the correspondent pairwise correlation vary with a number of relevant factors such the firing rate (de la Rocha et al., 2007). It is well known that correlations in pairs of neurons that fire few spikes per trial are weaker than in pairs that respond more strongly (Cohen and Kohn, 2011).\\
 
The aim of the present work is to develop statistical tools to study the synchronization strength between pairs of neurons during low firing rate conditions and, also, to discern subtle differences in synchronization during tonic activity.  There exist in the literature diverse methods to measure neural synchrony.  Harrison, et al. 2013 present a review on synchrony identification methods, including cross-correlation, joint peri-stimulus time histograms, trial-to-trial variability models, inhomogeneous Poisson models and generalized regression models for synchronous activity. Other include, the \textit{unitary-event analysis} (Gr\"un, 1996; Gr\"un, Diesmann, and Aertsen, 2002a,b), the \textit{conditional synchrony measure}, proposed in Faes et al., (2008), the \textit{event synchronization} method, proposed in Quian Quiroga, Kreuz and Grassberg, (2002) or the \textit{cross-nearest spike interval based method} proposed in Gonz\'alez-Montoro et al., (2014). In this occasion, we propose a cross-corre\-la\-tion based method referred to as  cross-corre\-la\-tion synchrony index (CCSI), in which the level of synchronization is measured as the area under the cross-correlation function in a neighborhood of zero with a correction for spurious coincidences.  In general, synchrony measures are based in binned spike trains. When firing rates are low, bins need to be defined large enough in order to decrease the number of zeros in the sequence. This procedure derives in the loss of important information, the timing of events. The measure we use estimates the distribution of the distances between spikes of different neurons in a smooth fashion, making a more efficient use of the information. We also propose  bootstrap methods for hypothesis testing. Resampling and bootstrap methodology is being widely used to assess for statistical significance of spike correlations (Gr\"un, 2009) and to conduct hypothesis tests in neuroscience in general (Kass, Ventura, and Brown, 2005). A fundamental approach to synchrony in spike trains is the jitter method (Amarasingham et al., 2012). However, the jitter procedure, together with other resampling methodology, are thought to assess synchrony against background variability. In the present work, special resampling methods are developed in order to imitate the existing associations between spike trains and conduct the bootstrap tests and, therefore, be able to test for differences in synchrony strength.\\

\section{Materials and methods}
\label{sec:2}

\subsection{Surgery and recording}\label{sec:2-1}
Experiments were carried out in adult anesthetized (iso\-flu\-ra\-ne) and paralyzed (gallamine) cats. The end-tidal CO2, body temperature and heart rate were continuously monitored and maintained under stable conditions. The animals were suspended on a stereotaxic frame and four craniotomies were performed in order to insert the recording and stimulating electrodes. All the procedures were performed by the researchers of  NEUROcom group and according to national and international guidelines.\\

Two recording methods were used: 1) in order to assess the global cortical activity, an electrocorticogram (ECoG) was obtained through a bipolar concentric electrode located in the primary somatosensory cortex; and, 2) an eight-channel multielectrode (FHC Inc.) was introduced into the primary visual cortex of the animal in order to make simultaneous extracellular recordings of several neurons. A guide-tube was used to arrange the tungsten electrodes in two rows and four columns, obtaining a grid of 8 recording points separated 200 microns and with independent vertical movement. Concurrently, two bipolar electrodes (tips separated 500 microns) were introduced for electrical stimulation at \textit{bs} and \textit{bf}.\\

Using the multielectrode we recorded spontaneous sleep-like activity of groups of neurons (1--3 units by electrode) for approximately 2 minutes. Next, \textit{bs} or \textit{bf} was stimulated for 2--4\,s (pulse trains of 2--8\,V, up to 1.5\,mA, and 50\,Hz) to induce an awake-like state lasting for up to 20\,s, followed by an spontaneous and slow return to the sleep-like state. The whole procedure was repeated several times for each activating pathway. In the present work each of these recordings is called a trial. The ECoG, spike waveforms and time stamps were conveniently filtered, digitized and stored for posterior analysis (Plexon Inc, Dallas, TX). Spikes were isolated using a box sorter during the experiment and re-sorted manually using the principal components of their waveforms (Offline Sorter, Plexon, Inc). Thus, to define a single unit, first and second principal components had to be (visually) well separated from the projections of all other waveforms (and noise) recorded on the same electrode.

\subsection{Cross-corre\-la\-tion synchrony index}\label{sec:2-2}

Let us consider two simultaneous spike trains, recorded in a $v$ seconds long time window around time $t$. Let us denote these spike trains by ${\cal X}=\{X_i\}_{i=1}^{m}$, $t-\frac{v}{2}<X_1<X_2<\cdots<X_{m}<t+\frac{v}{2}$, which we also call train 1,  and ${\cal Y}=\{Y_j\}_{j=1}^{n},$ $t-\frac{v}{2}<Y_1<Y_2<\cdots<Y_{n}<t+\frac{v}{2}$, also referred to as train 2. These processes need not to be Poisson processes. The only assumption we make is that the bivariate counting process associated to the simultaneously recorded spike trains, ${\cal X}$ and ${\cal Y}$, is stationary in the time period previous to the stimulus. We define $\lambda$ as the probability that, given a spike in train ${\cal X}$, there is a synchronous spike in train ${\cal Y}$. By \textit{synchronous} we mean that the two spikes occurred at the same time (or approximately the same time, with, at most, a difference of $\delta$) due to the synchronous activity between the two neurons and not to chance. Let us analogously define $\mu$ as the probability that, given a spike of train ${\cal Y}$, there is a synchronous spike in train ${\cal X}$. The synchrony measure we propose is the geometric mean of these two probabilities: 
\[{\cal T}= \sqrt{\lambda\mu}\]
which is a joint measure of the probability of synchronous firing. Note that, overall, ${\cal T}$  is a standardized measure of the joint firing rate.

\subsection{Estimation of ${\cal T}$}\label{sec:2-3}

Let us denote by $f$ the normalized cross-correlation function for lags up to $\frac{w}{2}$ seconds. Namely, $f$ is the density function of the distances between two spikes, one of each train, provided that that distance is smaller, in absolute value, than $\frac{w}{2}$. We call $w/2$ the \textit{cross-correlation window}.  To estimate ${\cal T}$ we propose to approximate the area under $f$ in a neighborhood of zero while taking into account the amount of that area that is due to chance. Let, 
\[A_{\delta}(\tau) = \int_{\tau-\delta}^{\tau+\delta} f(x) \textrm{dx}\]
and, to consider how much of $A_{\delta}(0)$ corresponds to chance, we propose to subtract the following approximation of the average of $A_{\delta}(\tau)$, for $\tau \in [\frac{-w}{2}, \frac{w}{2}]$, under no synchronous firing:

\begin{equation}\label{fact_corr}
\frac{1}{w} \int_{-\frac{w}{2}}^{\frac{w}{2}} A_{\delta}(y) \textrm{dy} = \frac{1}{w}\int_{-\frac{w}{2}}^{\frac{w}{2}}\int_{y-\delta}^{y+\delta}  f(x) \textrm{dx dy} \approx \frac{2\delta}{w}.
\end{equation}

The proof of (\ref{fact_corr}) can be found in Appendix A. Observe that both $\delta$ and $w$ are chosen regarding the problem and $f$ can be easily estimated with a normalized cross-correlogram or a kernel-smoothed version of it, which we denote by $\hat{f}$. Therefore, $A_{\delta}$ can be estimated as the corresponding integral of $\hat{f}$, which we denote by $\hat{A}_{\delta}$. So, to begin with,  we can define a synchrony measure as
\[\tilde{A}_{\delta} = \max{\{0, \hat{A}_{\delta}(0) - \frac{2\delta}{w}\}}\]

However, we aim to estimate $\sqrt{\lambda\mu}$. Let us study $\tilde{A}_{\delta}$ in a little bit more detail. $\tilde{A}_{\delta}$ describes the probability that, given two spikes, one of each spike train, they are synchronous. This last probability is actually the expected proportion of synchronous pairs of spikes among all possible pairs of spikes observed in $(t-\frac{v}{2}, t+\frac{v}{2})$ that are distant in less than $\frac{w}{2}$. Let 
$D=\{d_k\}_{k=1}^N=$ $\{X_i-Y_j: |X_i-Y_j|<\frac{w}{2}, i=1,\ldots,m, j=1,\ldots,n\}$ be the set of all possible differences between the spike times of one train and the spike times of the second one, which are, in absolute value, smaller than the cross-correlation window $\frac{w}{2}$. So, as just mentioned, $\tilde{A}_{\delta}$ estimates the expected proportion:

\begin{equation}\label{A_tilde}
 \frac{\#\{d_k \in D :  d_k \,\,\textrm{is due to synchrony}\} }{\#D}.
\end{equation}
Now, for every spike of train 1, the probability of finding a synchronous spike in train 2 was defined as $\lambda$, therefore, the expected number of synchronous pairs is $m\lambda$. On the other hand, for every spike of train 2, the probability of finding a synchronous spike in train 1 was defined as $\mu$ and, therefore, the expected number of synchronous pairs is $n\mu$. These two last mentioned expected values should be equal and, consequently, 
\[m\lambda = n\mu = \sqrt{m\lambda}\sqrt{n\mu}=\sqrt{mn}\sqrt{\lambda\mu}= \sqrt{mn}{\cal T}\]
which is the expected value of the numerator in (\ref{A_tilde}).\\

On the other hand,  the denominator in (\ref{A_tilde}) is the expected total amount of pairs that differ in less than $\frac{w}{2}$. For every spike observed in spike train 1, the expected number of spikes of train 2 that are closer than $\frac{w}{2}$ is $(n/v)w$ as $(n/v)$ is the mean firing rate of train 2 in the window $(t-\frac{v}{2}, t+\frac{v}{2})$. Then, in total,  $m(n/v)w$ is the expected number of pairs of spikes that differ in less than $\frac{w}{2}$.\\

Hence, $\tilde{A}_{\delta}$ is an estimator of
\[\frac{\sqrt{mn}\sqrt{\lambda\mu}}{m(n/v)w} = \frac{\sqrt{\lambda\mu}}{\sqrt{nm}(w/v)}.\]

This last expression suggest the definition of the estimator $\widehat{\cal T}_{\delta}$ of ${\cal T}$, at time $t$, as some corrected version of $\tilde{A}_{\delta}$, that we will refer to as the cross-correlation synchrony index (CCSI):

\begin{equation}\label{CCSI}
\widetilde{\cal T}_{\delta} = \tilde{A}_{\delta}\sqrt{nm}(w/v)
\end{equation}

As already stated, we are interested in the time evolution of synchrony during the spontaneous sleep-like oscillatory activity and also during the awake-like tonic activity induced after electrical stimulation.  Therefore, in order to estimate time-varying synchrony we use moving windows of size $v$ and estimate ${\cal T}$ in each window, obtaining $\widetilde{{\cal T}}_{\delta}(t)$.\\

\subsection{Nonparametric smoothing of $\widetilde{\cal T}_{\delta}$}\label{sec:2-4}

The number of spikes at each time window, $V_t$, is very variable and when this number is small $\widetilde{\cal T}_{\delta}(t)$ becomes less reliable. To make the CCSI more robust, in order to be able to highlight characteristics of these curves and find patterns due to experimental conditions, we use a kernel smoother of the form:
\[
\widehat{\cal T}_{\delta}(t)=\sum_{j=1}^M \Psi_j(t)\widetilde{\cal T}_{\delta}(t)
\]
for some weight functions $\Psi_j$. The most common kernel estimator is the Nadaraya-Watson estimator (Nadaraya, 1964; Watson, 1964), which is constructed using the weights

\[\Psi_j(t)=\frac{\textsl{K}_h(t_j-t)}{\sum_r \textsl{K}_h(t_r-t)}\]
with $\textsl{K}_h(u)=\frac{1}{h}\textit{K}(\frac{u}{h})$ and $\textsl{K}$ a suitable kernel function. For our data, we used the uniform kernel function $\textsl{K}(u)=0.5$ if $|u|<1$ and $0$ otherwise and the smoothing parameter $h=5$.\\

\subsection{Testing for Synchrony Differences}\label{sec:2-5}

To check whether there are differences between the CCSI during the awake-like period and the CCSI during sleep-like spontaneous activity, a hypothesis test is implemented. As a result of the assumption that the bivariate process is stationary before the stimulus onset, the synchrony is constant during that time period, i.e., ${\cal T}(t)={\cal T}_0$  for every $t\in [0,t_{st})$, where $t_{st}$ denotes the time point where stimulation was applied. However, the synchrony needs not to be constant after the appearance of the stimulus. Therefore, we aim to test whether the synchrony at two given points, $t_0\in [0,t_{st})$ and $t_1\in (t_{st}, T]$, is equal, or, equivalently, whether at time $t_1\in (t_{st}, T]$ (awake-like activity, after the stimulus) the synchrony index has recovered its stationary sleep-like value. In this context, the null ($H_0$) and alternative ($H_1$) hypotheses can be stated as follows. For some fixed $t\in(t_{st},T]$:

\begin{itemize}
\item[] $H_0: {\cal T}(t)={\cal T}_0$  
\item[] $H_1: {\cal T}(t)<{\cal T}_0$.
\end{itemize}

The stationarity of the bivariate counting process in $[0,t_{st}]$ implies that the sampling distribution (before the stimulus) of the joint trains $(\cal X,{\cal Y})$  does not depend on $t$. As a consequence, ${\cal T}(t)$, and the sampling distribution of $\widehat{\cal T}_{\delta}(t)$ do not depend on $t$ in the period previous to the stimulus. An important issue is, therefore, to approximate the probability distribution of the test statistic $\widehat{\cal T}_{\delta}(t)$ under the null hypothesis. We propose an extension of the stationary bootstrap (Politis and Romano, 1994) to imitate the  synchronous activity in the sleep-like time interval. Politis and Romano's stationary bootstrap is a resampling mechanism that mimics the underlying distribution of the stochastic process and also preserves stationarity. These two properties are preserved by our extension to a bivariate setting. Since, stationarity implies constant synchrony, using the stationary bootstrap guarantees that the resampling distribution of \mbox{$\widehat{\cal T}^*_{\delta}(t)$}, mimics the sampling distribution of \mbox{$\widehat{\cal T}_{\delta}(t)$} under the null hypothesis, this is \mbox{${\cal T}(t) = {\cal T}_0$}. The proposed procedure produces two synchronous bootstrap spike trains,  which are then used to calculate bootstrap versions of \mbox{$\widehat{\cal T}_{\delta}(t)$} under $H_0$. The procedure can be described as follows:\\

\begin{enumerate}
	\item Merge the two trains into one but take note of which spike belonged to which original train, this is, label them. 
	\item Compute the Inter Spike Intervals (ISI) of this joint train. Let us call this sequence of values $\mathbf{S}$.
	\item Build $\mathbf{S}^1$ and $\mathbf{S}^2$ the sets of ISI that start in a spike of neuron 1 and 2 respectively.
	\item Choose at random one ISI from $\mathbf{S}$. 
	\item Choose the following value in $\mathbf{S}$ with probability $1-p_{boot}$. With probability $p_{boot}$ choose an ISI from the corresponding $\mathbf{S}^i$. This is, if the last spike came from spike train 1 then the next ISI needs to be chosen from $\mathbf{S}^1$ and from $\mathbf{S}^2$ in the other case. 
	\item Repeat Step 5 until obtaining enough resamples time to imitate the true sampling time.
	\item Use the labels on the ISI to separate the obtained train into two bootstrap trains.
	\item Compute the CCSI for the bootstrap pair of spike trains. 
	\item Repeat  Steps 4--8 $B$ times to obtain $B$ bootstrap replicates of the CCSI.     
\end{enumerate}

This algorithm simulates the joint distribution of the bivariate counting process under the null hypothesis and therefore reproduces the distribution of $\widehat{\cal T}_{\delta}(t)$ under the null. To facilitate the reading we delay a more technical description of the algorithm to Appendix B. \\

We seek for significant reductions in synchronization during the awake-like period with respect to the sleep-like period, so we computed a significance threshold as follows. For each $b=1,\cdots,B$ and each time $t$, $\widehat{\cal T}_{\delta}^{*b}(t)$ is a bootstrap analogue of $\widehat{\cal T}_{\delta}(t)$ under $H_0$. We assume  constant synchrony in the period preceding the stimulus onset. Therefore, the $\alpha$-quantile of the set of all values of the bootstrap curves is a plausible choice for a critical value. We denote this value by $\widehat{\cal T}^{*}_{\alpha}$. The null hypothesis is rejected at each time point $t\in(t_{st},T]$ if $\widehat{\cal T}_{\delta}(t)<\widehat{\cal T}^{*}_{\alpha}$. Observe that these choice of the critical value takes into account multiple testing.

\subsection{Testing for the difference between the two activating pathways}\label{sec:2-6}

Apart from differences in synchrony profiles before (sleep-like) and after (awake-like) the stimulus onset, we aim to develop a method to study differences in synchronization dynamics during the awake-like period induced by the activation of the two different pathways. In this context the relevant hypotheses are:
\begin{itemize}
\item[] $H_0: {\cal T}^{\textit{bs}}(t)= {\cal T}^{\textit{bf}}(t)$ for all $t \in [t_{st}, T]$
\item[] $H_1: {\cal T}^{\textit{bs}}(t)\neq {\cal T}^{\textit{bf}}(t)$ for some $t \in [t_{st}, T]$.
\end{itemize}
where ${\cal T}^{\textit{bs}}(t)$ and ${\cal T}^{\textit{bf}}(t)$ are ${\cal T}$ under \textit{bs} or \textit{bf} stimulation respectively. We develop a test that enables to detect in which time periods there are differences, if any, between the awake-like activity induced by each pathway. The test statistic we use is 
$$T_{CCSI}(t)=\widehat{\cal T}_{\delta}^{\textit{bs}}(t)-\widehat{\cal T}_{\delta}^{\textit{bf}}(t).$$
In this case, due to the expected lack of stationarity induced by the stimulation, the bootstrap setup used in the previous section is not valid. As the test is to be applied in the time interval after stimulation, we make use of the different trials for resampling. Roughly speaking, the method consists in shuffling the trials but taking into account the temporal dependence. This is, blocks of random lengths are chosen from randomly chosen trials but the order (in time) of the blocks is respected:

\begin{enumerate}
	\item Merge each trial of the pair of neurons into one train. This is, obtain one merged train for each recorded trial and keep the information of  what original train each spike corresponds to. 
	\item Choose one trial at random and choose the first spike of that merged trial. 
	\item Given an already chosen spike time, $x^*_i = x_j$, choose the next spike as $x^*_{i+1} = x_{j+1}$ from the same trial with probability $1-p_{boot}$. With probability $p_{boot}$, choose another trial, say $k$, and define $x_{i+1}^*$ (the next spike time for the bootstrap merged train)  as the next spike time in trial $k$.
	\item Repeat this procedure to obtain as many resampled trials as in the real data scenario.
	\item Separate the resampled trials into the corresponding bootstrap trains. 
	\item Compute the bootstrap CCSI for each stimulus using the bootstrap trials.
	\item Repeat Steps 4--6 $B$ times to obtain $B$ replicates of the CCSI for each stimulus and compute the difference between the two. 
	\item Compute the desired quantiles to reject the null hypothesis if the observed difference is outside the interval defined by them.
	
\end{enumerate}

If $H_0$ holds, the trials for both stimuli are generated by the same process. The proposed bootstrap mimics that process using the pooled information in Steps 1--4 above. As before, a detailed description of the algorithm can be found in Appendix B.\\

\section{Results}\label{sec:3}

The performance of the method and bootstrap tests was examined in real data from the experiment described in Section~\ref{sec:2-1}. We used three simultaneously recorded neurons: A, B and C grouped in two pairs: A-B and A-C.\\

\begin{figure}[h!]
\centering
\includegraphics[scale=0.5]{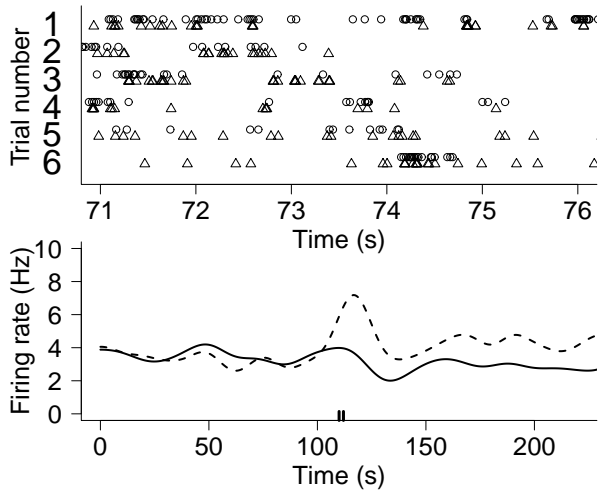}
\caption{Top panel: Raster plots of 5 seconds of simultaneous recordings of spikes of neuron A (circles) and B (triangles). Bottom panel: Firing rates of neurons A (solid line) and B (dashed line) averaged over three trials for each neuron.}\label{fig:firingrates}
\end{figure}

As the data come from spontaneous activity recordings, the firing rates are very low for each neuron and, as expected, it is very hard to find spikes occurring at exactly the same instant. The top panel of Figure~\ref{fig:firingrates} shows five seconds of the simultaneous recording of two neurons during sleep-like spontaneous activity. Spikes of neuron A are represented by circles and spikes of neuron B by triangles. The low frequency of exact matches can be observed. However, it is clear that neuron spikes are not independent and that there exists synchrony up to some extent. The low firing rate present in all the recording time can be easily observed in the bottom panel of Figure~\ref{fig:firingrates} (the instant of stimulation is indicated at time 110~s). The firing rates were estimated using kernel smoothers with a Gaussian kernel function and a bandwidth selected ad hoc for illustrative purposes.\\

Under these experimental conditions, we considered synchrony as the event of two neurons firing together up to a time lag of $\delta=0.025$\,s. A cross-corre\-la\-tion window of $2\nu=2$\,s was also used. Also, $\alpha=0.05$ has been considered for all the analyses. To estimate CCSI at time $t$ we used the activity in a neighborhood of 10\,s around $t$.\\ 

\begin{figure}[h!]
\centering
\includegraphics[scale=0.45]{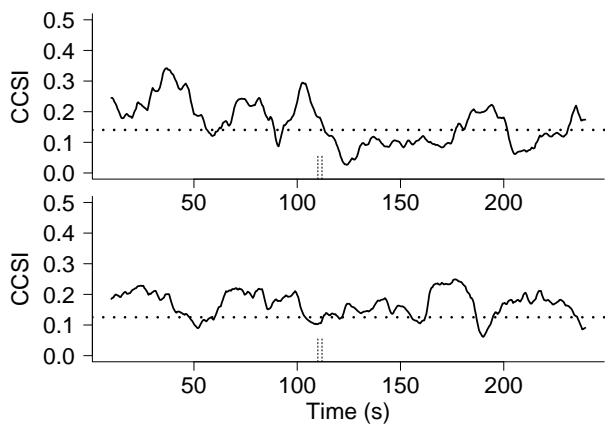}
\caption{Estimated CCSIs (solid lines) averaged over the three trials of the pair A-B and significance threshold (horizontal dotted line) for differences in synchrony estimated with the bootstrap procedure described in Section \ref{sec:2-5}. The period of stimulation is indicated by the vertical dashed lines on the x-axis. Top panel: \textit{bs}. Bottom panel: \textit{bf}.} \label{fig:boot11}
\end{figure}

\begin{figure}[h!]
\centering
\includegraphics[scale=0.45]{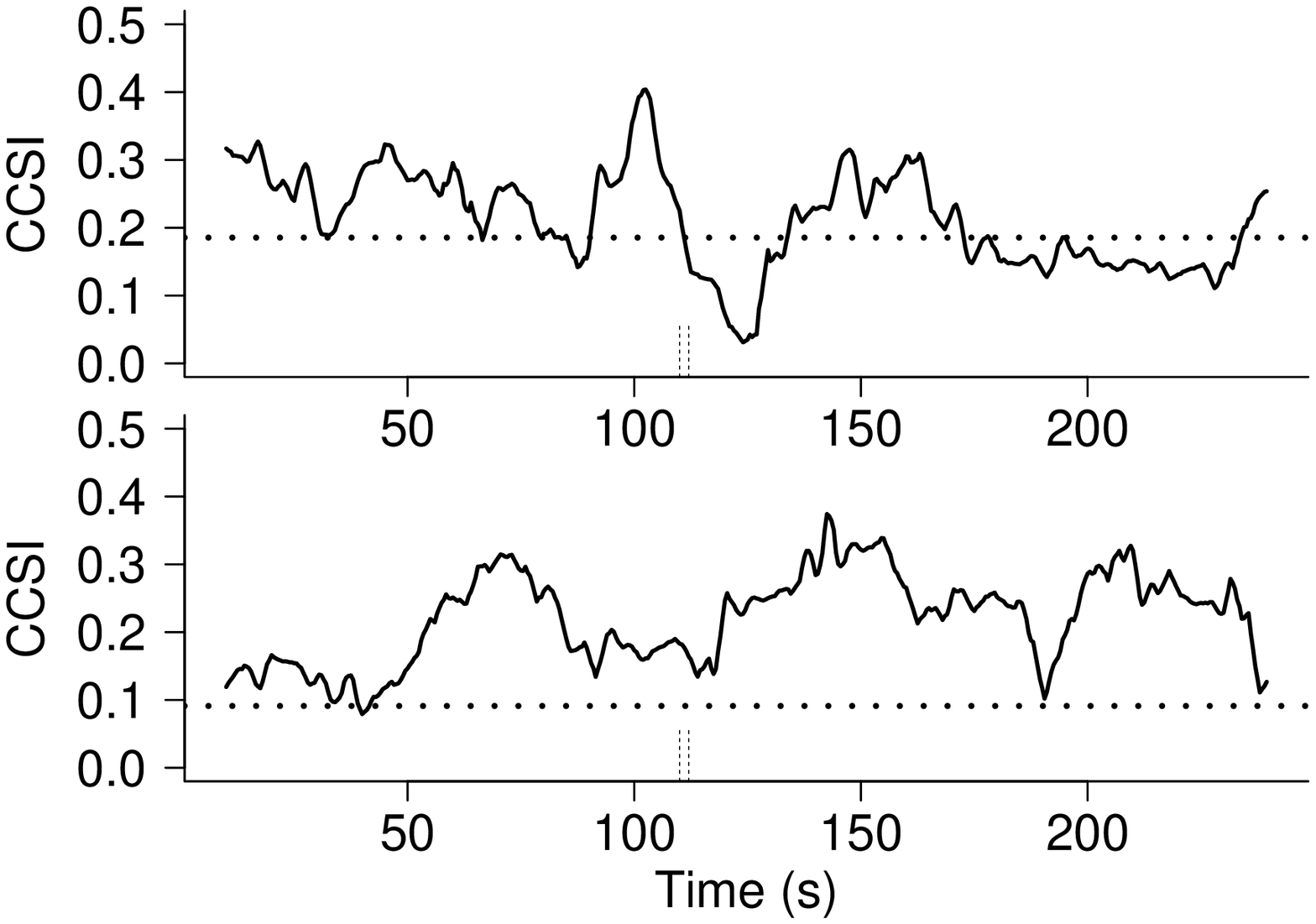}
\caption{Estimated CCSIs (solid lines) averaged over the three trials of the pair A-C and significance threshold (horizontal dotted lines) for differences in synchrony estimated with the bootstrap procedure described in Section \ref{sec:2-5}. The period of stimulation is indicated by the vertical dashed lines on the x-axis. Top panel: \textit{bs}. Bottom panel: \textit{bf}.} \label{fig:boot12}
\end{figure}

To test for stimuli effects on synchrony over time, the bootstrap procedure described in Section \ref{sec:2-5} was used with parameter $p_{boot}=0.01$. The choice of the bootstrap parameter was made on an attempt to reach a balance between imitating the dependence in the data and the variability the method pursues. Having in mind that our curves are sampled in 460 points and noting that the lengths of the resampled data blocks follow a geometric distribution, this choice of $p_{boot}$ results in resampling, in the mean,  5 blocks of length 92. Fi\-gu\-re~\ref{fig:boot11} shows the results for the existing synchrony between neurons A and B and Figure~\ref{fig:boot12} shows the same results for the pair A and C. In these examples we can  detect subtle changes in the synchronization dynamics for both pairs under the \textit{bs} stimulation. In this case, a significant decrease of synchrony immediately after stimulation can be observed. \\

\begin{figure}[h!]
\centering
\includegraphics[scale=0.42]{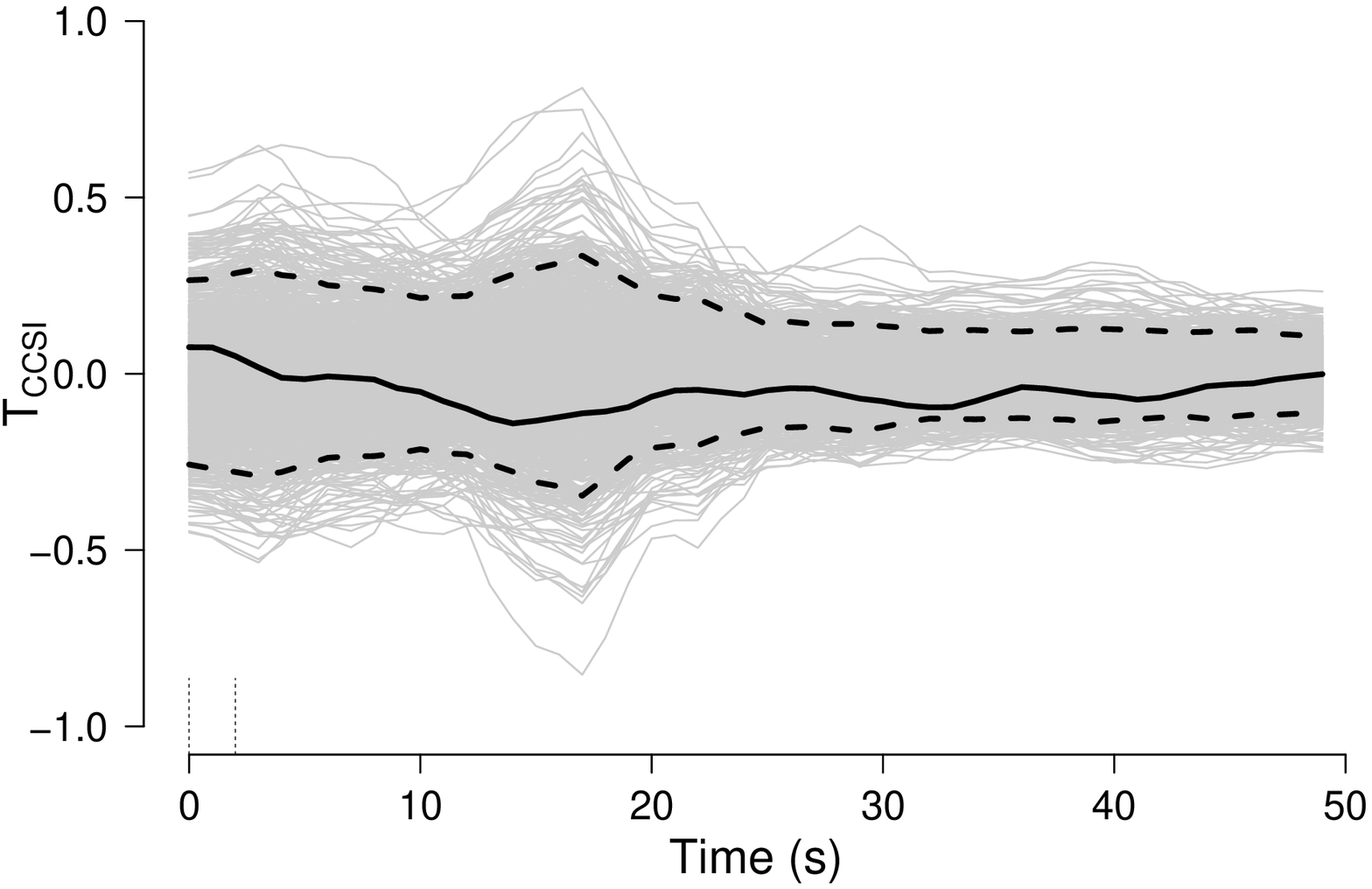}
\hspace{2cm}
\includegraphics[scale=0.42]{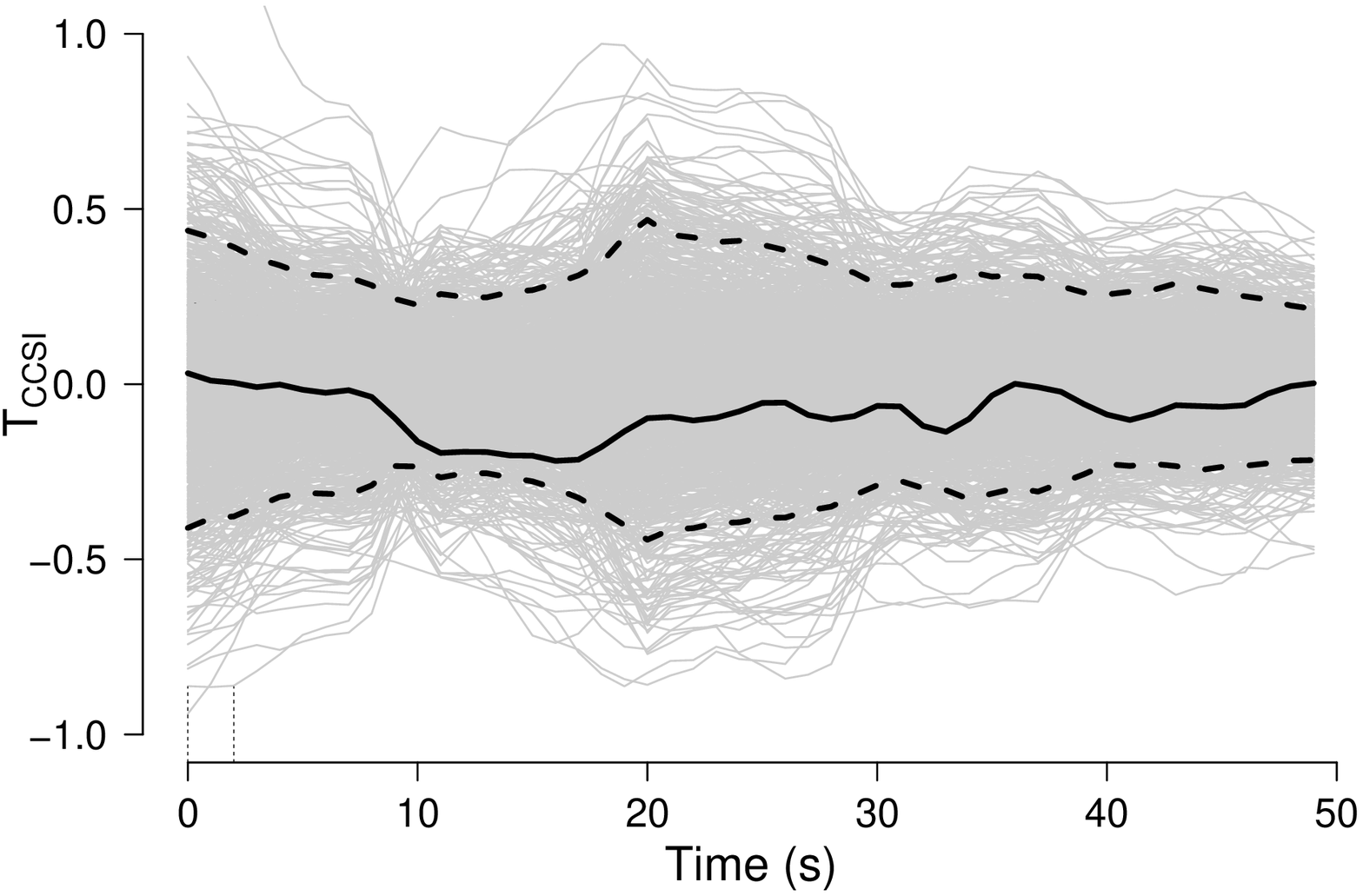}
\caption{Differences of CCSI (solid black lines) with confidence bands (dashed black lines) built with 1000 bootstrap replications (grey lines).  The stimulus is shown for time $[0,2]$ using vertical lines on the horizontal axis. Top panel: pair A-B. Bottom panel: pair A-C.} \label{fig:boot2}
\end{figure}

Figure~\ref{fig:boot2} shows the results obtained when testing for the effect of stimulation in the difference between the $\widehat{\cal T}_{\delta}$ obtained under each stimulus. Different $\widehat{\cal T}_{\delta}$ curves are shown together with $95\%$ confidence bands obtained using the bootstrap procedure described in Section~\ref{sec:2-6}. The \textit{difference curves} for both pairs are shown as solid black lines. Although they deviate from zero,  none of them result significant. It can be observed that the bootstrap confidence bands result very wide, this is probably a consequence of the small number of trials that are available (needed for the resampling)  and the high trial to trial variability.\\

\section{Simulation study}\label{sec:4}

\subsection{Power of the test}\label{sec:4-1}

A simulation study was carried out to show the performance of the method. For this aim, we simulated pairs of spike trains controlling their association.  We used an underlying Poisson process with rate $\lambda(t)$, say $M^0(t)$. To generate two spike trains from this underlying process, we assumed to  have a realization of $M^0(t)$ with events at $X^0_1, \ldots, X^0_N$ and two vectors of random errors $\mu^1=(\mu^1_1, \ldots, \mu^1_N)$ and  $\mu^2=(\mu^2_1, \ldots, \mu^2_N)$ with $\mu^j_i$ sampled from a uniform distribution chosen accordingly to the firing rate as explained below.  Let $M^1(t)$ and $M^2(t)$ be the pair of spike trains induced by $M^0(t)$ as follows:

\[
	P(M^j(t)-M^j(t^-)=1)=\left\{\begin{array}{lc}
	p^j(t)&\textrm{\small{if}}\,\, t=X_i^0+\mu_i^j\\
	& \textrm{\small{for some}}\, i=1,... , N\\
	0& \textrm{\small{otherwise}}\end{array}\right.
\]
for $j=1,2$ and $p^j(t)$ a certain probability function defined for each train.\\

In order to introduce changes in synchrony, we considered a time point, $t_0$, as the time where the association between the trains change. So, the probabilities are set constant before $t_0$ and also constant (although with a different value) after $t_0$. Also, for simplicity, we used the same probabilities for both trains.  This is:

\[
	p^1(t)=p^2(t)=\left\{\begin{array}{lc}
	p_1&\textrm{if}\,\, t<t_0\\
	p_2&\textrm{if}\,\, t\geq t_0 \end{array}\right..
\]
On the other hand, we defined the firing rate of the trains as constant throughout the trial, say $\lambda_0$.  Therefore, the firing rate of the process $M^0(t)$ is defined as

\[
	\lambda(t)=\left\{\begin{array}{lc}
	\lambda_0/ p_1 &\textrm{if}\quad t<t_0\\
	\lambda_0/ p_2 &\textrm{if}\quad t\geq t_0 \end{array}\right..
\]
In practice, we drew  random numbers $\rho^j_i\in [0,1]$ and then selected $X_i^0+\mu_i^j$ as a spike for train $j$ if $\rho^j_i\leq p^j_i$ (which occurs with probability $p_i^j$).\\

Finally, for the simulation study, $220$\,s spike trains with constant rate of $\lambda_0=4$\,Hz were generated: $110$\,s were simulated with probability $p_1$ of acquiring the spikes from the underlying process and another $110$\,s with probability $p_2$. We used $\mu^j_i\sim\textit{U}(-1/(20\lambda_0), 1/(20\lambda_0))$ for all $i=1,\ldots, N$ and $j=1,2$, in order to have a controlled error which shifts the spikes in a small amount but so that it is not likely that one spike would be shifted so much that would get very close to  another spike.   The choices for the parameters $p_1$ and $p_2$ were $p_1=0.7$ with  $p_2=0.1,0.3,0.5 $ and $0.65$ on the other. We simulated 500 pairs of trains and estimated the CCSI function from them using the same parameters as for the real data ($\delta\nu=0.025$\,ms, $v=10$\,s.). Then we performed the bootstrap test described in Section~\ref{sec:2-5} with $B=500$ and $p_{boot}=0.01$ the same as for the real data.\\

\begin{figure}[h!]
\centering
\includegraphics[scale=0.45]{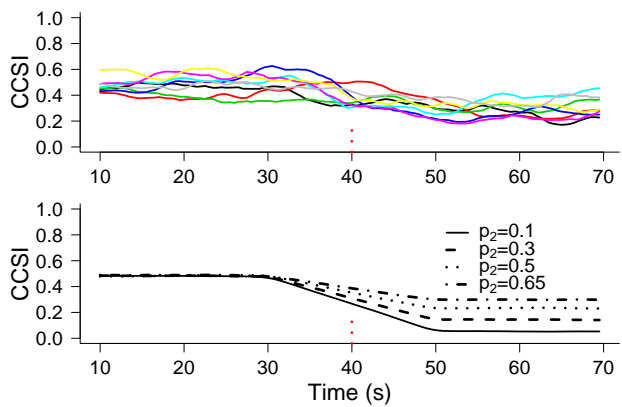}
\caption{$CCSI$  curves for eight simulated pairs of neurons using $p_1=0.7$ and $p_2=0.65$ (top panel) and average of 500 $CCSI$ curves for $p_1=0.7$ and $p_2=0.1$ (solid line), $0.3$ (dashed line), $0.5$ (dotted line) and $0.65$ (dashed-dotted line) (bottom panel). The stimulus is simulated at $t=40$ (vertical red dotted line).} \label{fig:sim0}
\end{figure}

\begin{figure}[h!]
\centering
\includegraphics[scale=0.45]{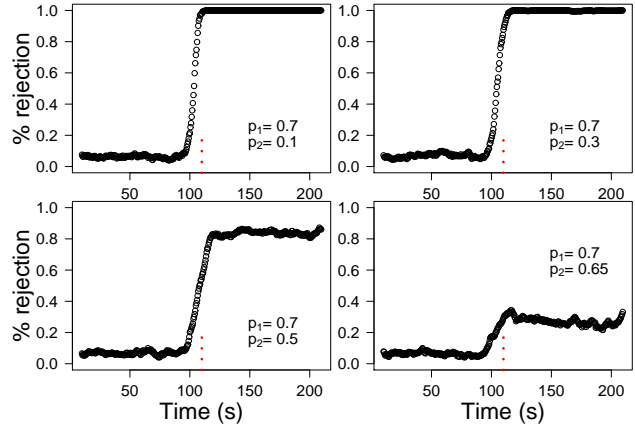}
\caption{Rejection proportion of the bootstrap test for changes in synchrony using $p_1=0.7$ and $p_2= 0.1, 0.3, 0.5$ and $0.65$. The stimulus is simulated at $t=40$.} \label{fig:sim1}
\end{figure}

Figure~\ref{fig:sim0} shows eight CCSI curves from eight pairs of simulated spike trains with $p_1=0.7$ and $p_2=0.65$ in the top panel, whereas in the bottom panel the average of 500 of these curves are shown for different choices of $p_2$.  Figure \ref{fig:sim1}  shows the rejection percentage of the null hypothesis using the bootstrap test presented in Section~\ref{sec:2-5} with $p_1=0.7$ and different values for $p_2$.  As it can be observed the test can easily detect the changes in synchrony. Of course the use of sliding windows provokes the existence of a period of time where the rejection percentage grows slowly. This is one of the prices we have to pay because of the low firing rates.\\

 The rejection percentage before stimulation in Figure~\ref{fig:sim1}, is the percentage of rejection under the null hypothesis, this is, the level of the test. Although the values do not  reach the nominal level ($0.05$)  the results are acceptable, given the difficulty of the problem due to the low firing rates. In average, over all windows of all simulations, the level is $0.065$. Also, in simulations with higher firing rates ($10 Hz$) we reached a level of $0.053$.
\\

On the other hand, Figure~\ref{fig:sim1} shows very good results for the power of the test (except in the case where $p_2 =0.65$). As expected, it decreases with the shortening of the difference in amount of synchrony between the pre and post-stimulus intervals. In all simulations, the probability of joint-firing in the pre-stimulus part is 0.7. When the probability of joint-firing in the post part is small ($p_2=0.1$) the power is 1. When we increase the post joint-firing probability to 0.3 the power decreases to 0.998. If we consider the post joint-firing probability to be 0.5, the power is $0.83$. Finally, if the joint-firing probability in the post part is considered to be 0.065 (almost no difference with the pre-stimulus part) the power decreases to $0.26$. This values are averages over the time window  $[120,200]$.

\subsection{Firing rate effect}\label{sec:4-2}
\begin{figure}[h!]
\centering
\includegraphics[scale=0.4]{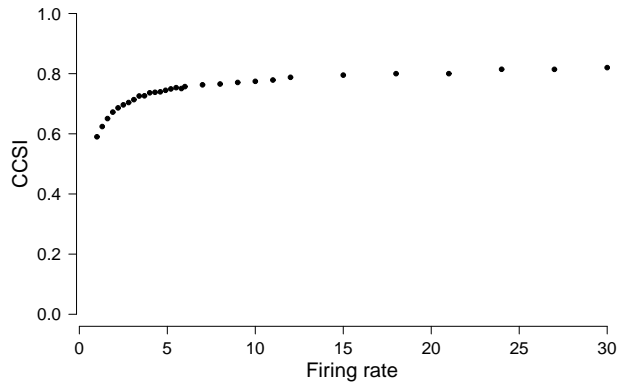}
\caption{Average of 1000 CCSI values obtained at different firing rate values.} \label{fig:rates}
\end{figure}

With the same simulation procedure, we performed a simulation study to evaluate how the firing rate affects the performance of the index. For several firing rate values we simulated 1000 twenty seconds long spike trains and computed the CCSI for each one. Figure~\ref{fig:rates} shows the average of the 1000 obtained values for each firing rate value. The fluctuation of CCSI is extremely low for moderate or large firing rates (8--30~Hz). For small firing rates (1--8~Hz) the fluctuation of CCSI is moderate (0.6--0.75). As a consequence, having in mind the firing rates in Figure~\ref{fig:firingrates}, the practical influence of the firing rate in CCSI is low. 

\section{Discussion}\label{sec:5}

We have used a cross-corre\-la\-tion based synchrony index to study pairwise synchrony between  cortical neurons under spontaneous activity. The method is based on kernel estimation of the cross-corre\-la\-tion function and its integration in a neighborhood around zero.  It is a flexible method because it permits the tuning of its parameters to better fit the problem. Two hy\-po\-the\-sis tests have been proposed to test for differences in synchrony profiles. Resampling methods are very useful and powerful tools when no parametric model can be assumed for the data, as often occurs with spike activity. The first proposed resampling procedure takes into account the dependence between simultaneous spike trains by resampling from the intervals of time that elapses between spikes of a joint spike train built by merging the spike trains. The second one, takes into account the dependence by shuffling trials but respecting the timing of spikes. The methods have been used in real data to test for synchrony dynamics between neurons under two different induced states (sleep-like and awake-like) and after the activation of two ascending pathways (brainstem and basal forebrain).  The results show that the proposed method works efficiently under conditions of low neural activity. It is also shown that the method is useful to discriminate between subtle changes in synchronization dynamics. The good performance of the first bootstrap test has been assessed by a simulation study. Although the methods were thought for a particular problem, we  believe they  can be used in many other contexts, especially when low firing rates are an issue but not only then.

\section*{APPENDIX A}
Approximation to the average of $A_{\delta}(\tau)$ in (\ref{fact_corr}):

\begin{dmath*}
\frac{1}{w} \int_{-\frac{w}{2}}^{\frac{w}{2}} A_{\delta}(y) dy = \frac{1}{w} \int_{-\frac{w}{2}}^{\frac{w}{2}} \int_{y-\delta}^{y+\delta} f(x) dx dy= 
\frac{1}{w} \int_{-\frac{w}{2}-\delta}^{\frac{w}{2}+\delta} \int^{\min\{x+\delta, \frac{w}{2}\}}_{\max\{x-\delta, -\frac{w}{2}\}} f(x) dy dx=
\frac{1}{w} \int_{-\frac{w}{2}}^{\frac{w}{2}} (\min\{x+\delta, \frac{w}{2}\} - \max\{x-\delta, -\frac{w}{2}\}) f(x) dx=
\frac{1}{w} \left[ 2\delta\left(\int_{-\frac{w}{2}+\delta}^{\frac{w}{2}-\delta} f(x) dx\right)  + \int_{-\frac{w}{2}}^{-\frac{w}{2}+\delta} (x+\delta+\frac{w}{2})f(x) dx + \int_{\frac{w}{2}-\delta}^{\frac{w}{2}} f(x) (\frac{w}{2} - x +\delta) dx \right] = 
\frac{1}{w} \left[ 2\delta \int_{-\frac{w}{2}}^{\frac{w}{2}} f(x) dx - 2\delta \int_{\frac{w}{2}-\delta}^{\frac{w}{2}} f(x) dx - \left.
\right. 2\delta \int_{-\frac{w}{2}}^{-\frac{w}{2}+\delta} f(x) dx + \int_{-\frac{w}{2}}^{-\frac{w}{2}+\delta} (x + \delta + \frac{w}{2})f(x) dx + 
\int_{\frac{w}{2}-\delta}^{\frac{w}{2}} f(x) (\frac{w}{2} - x + \delta) dx\right] 
\end{dmath*}

\begin{dmath*}
\approx \frac{1}{w} \left[ 2\delta -2\delta\delta f(\frac{w}{2}) - 2\delta\delta f(-\frac{w}{2}) +
f(-\frac{w}{2}) \int_{-\frac{w}{2}}^{-\frac{w}{2} + \delta} (x + \delta + \frac{w}{2}) dx + 
f(\frac{w}{2}) \int_{\frac{w}{2}-\delta}^{\frac{w}{2}} (\frac{w}{2} - x + \delta) dx \right] = 
\frac{1}{w} \left[ 2\delta-2\delta^2 f(\frac{w}{2}) - 2\delta^2 f(-\frac{w}{2}) + f(-\frac{w}{2}) \frac{3\delta^2}{2} + 
f(\frac{w}{2}) \frac{3\delta^2}{2} \right] = 
\frac{1}{w} \left[2\delta + O(\delta^2)\right] \approx \frac{2\delta}{w}
\end{dmath*}

\section*{APPENDIX B}
The bootstrap procedure in Section~\ref{sec:2-5} is detailed here. 

\begin{enumerate}
\item[1.] Merge the two observed trains, ${\cal X}_1$ and ${\cal X}_2$, into one, ordering all the spiking times together in a joint train. Let the pooled train be \\${\cal X}^p=\{(X^c_1,\gamma^p_1),\ldots, (X^p_N,\gamma^p_N)\}$ where \\
$\gamma^p_i=\left\{\begin{array}{cc}
1 &\textrm{if}\,\, X_i^p\in {\cal X}_1\\
2 &\textrm{if}\,\, X_i^p\in {\cal X}_2\\
\end{array}\right.$

This is, $\gamma^p_i$  is an indicator variable of the spike train to which the action potential that occurs at time $X^p_i$ belongs.
\item[2.] Next,  compute the interspike intervals (ISI) of this new train: $S^p_1=X^p_1$ and $S^p_{i+1}=X^p_{i+1}-X^p_i, i=1,\ldots, N-1$ and let $\textbf{S}^p=\{(S^p_i, \gamma^p_i)\}_{i=1}^N$.
\item[3.] Build the sets\\
$\textbf{S}^1=\{(S^p_i, \gamma^p_i): \gamma^p_{i-1}=1; i=1,\ldots,N\}$ and $\textbf{S}^2=\{(S^p_i, \gamma^p_i): \gamma^p_{i-1}=2; i=1,\ldots,N\}$. This is, $\textbf{S}^1$ (and respectively  $\textbf{S}^2$) contains the elapsed times from a spike of neuron 1 (respectively 2) to the following spike in the joint train, and their corresponding neuron indicators.
\item[4.] Randomly choose $(S_1^{p*},\gamma^{p*}_1)$ from  $\textbf{S}^p$, i.e. \\$P^*((S_1^{p*},\gamma^{p*}_1)=(S_i^{p},\gamma^{p}_i))=\frac{1}{N}$ $i=1,\ldots,N$.
\item[5.] If $S^{p*}_i=S^p_j$ choose $(S_{i+1}^{p*},\gamma_{i+1}^{p*})=(S_{j+1}^{p},\gamma_{j+1}^{p})$ [in the case $j=N$, $(S^{p*}_{i+1},\gamma^{p*}_{i+1})=(S^{p}_{1},\gamma^p_1)$] with probability $1-p_{boot}$ and choose it at random from $\textbf{S}^{\gamma^p_j}$ with probability $p_{boot}$.
\item[6.] Repeat Step 5 until obtaining the first $(S_M^{p*},\gamma_M^{p*})$ for which $\sum_{i=1}^M S_i^{p*}\geq t_{st}$.
\item[7.] Build the ISIs for the first bootstrap train, ${\cal X}^{1*}$. Let $L_1=\min_{l}\{\gamma^{p*}_l=1\}$, then \mbox{$S^{1*}_1=\sum_{k=1}^{L_1}S_k^{p*}$}. For $i=2,\ldots, I_1=\#\{\gamma^{p*}_l:\gamma^{p*}_l=1\}$ let $L_i=\min_{l>L_{i-1}}\{\gamma^{p*}_l=1\}$,  and then $S^{1*}_{i}=\sum_{k=L_{i-1}+1}^{L_i}S^{p*}_k$.
\item[8.] Build the first bootstrap train ${\cal X}_1^{*}$ as  $X^{*}_{1i}=\sum_{k=1}^{i}S^{1*}_k$ for $i=1,\ldots,I_1$.
\item[9.] Build the second bootstrap train ${\cal X}^{*}_2$ in a similar way. This consists in repeating Steps 7--8 but with the condition $\gamma^{p*}_l=2$.
\item[10.] Based on ${\cal X}^*_1$ and ${\cal X}^*_2$ compute $\widehat{\cal T}_{\delta}^*(t)$ as in Subsections~\ref{sec:2-3} and \ref{sec:2-4}.
\item[11.] Repeat Steps 4--9 B times to calculate $\widehat{\cal T}_{\delta}^{*b}(t)$, $b=1,\ldots,B$, for these bootstrap trains.
\end{enumerate}

Steps 1--3 in the algorithm are used to build the joint train. Bootstrap resamples for the ISIs of this joint train are obtained in Steps 4--6. Finally Steps 7--9 separate the joint train to obtain two `simultaneously recorded' bootstrap trains. \\

Now we present the detailed algorithm for the bootstrap procedure introduced in Section~\ref{sec:2-6}.

\begin{enumerate}
\item[1.] Build a joint train for each recorded trial $k$, $k=1,\ldots, K$:\\
 ${\cal X}_k^p=\{(X_{k1}^p,\gamma_{k1}^p),\ldots, (X_{kN_k}^p,\gamma_{kN_k}^p)\}$ where, as above, 
\[\gamma^p_{ki}=\left\{\begin{array}{cc}
1 &\textrm{if}\,\, X^p_{ki}\in {\cal X}_1\\
2 &\textrm{if}\,\, X^p_{ki}\in {\cal X}_2\\
\end{array}\right.\] 
\item[2.] Choose a trial, $k_1$, at random with equal probability from $\{1,\ldots,K\}$ and define $X_1^{p*}=X^p_{k_11}$ and $\gamma_1^{p*}=\gamma_{k_11}^p$.
\item[3.] If  $(X_{i}^{p*}, \gamma_{i}^{p*})=(X_{k_ij}^{p}, \gamma_{k_ij}^{p})$ then, with probability $1-p_{boot}$ set $k_{i+1}=k_i$, $X_{i+1}^{p*}=X_{k_{i+1} (j+1)}^{p}$ and $\gamma_{i+1}^{p*}=\gamma_{k_{i+1} (j+1)}^{p}$. With probability $p_{boot}$, draw $k_{i+1}$ at random with equal probabilities from  $\{1,\ldots, K\}$, set $X_{i+1}^{p*}=X_{k_{i+1} m}^{p}$ so that $X_{k_{i+1}m}^{p}=\min_{l}\{X_{k_{i+1}l}^{p}>X_i^{p*}\}$ and $\gamma_{i+1}^{p*}=\gamma_{k_{i+1} m}^{p}$.
\item[4.] Increase the index $i$ by one unit and repeat Steps 2--3 while possible, i.e., while there exists some index $l$, such that $X_{k_{i+1}l}^p>X_i^{p*}$.
\item[5.] For each trial, $k$, and each stimulus, $j=1,2$, repeat Steps 2--4 above to obtain the bootstrap train ${\cal X}^*_{jk}$.
\item[6.] For each trial, $k$, separate the two spike trains using the information gathered by $\gamma^p$ to get $X_{1k}^*$ and $X_{2k}^*$.
\item[7.] Compute the bootstrap $\widehat{\cal T}_{\delta}^{s*}(t)$ for each stimulus, $s\in\{\textit{bs}, \textit{bf}\}$ and $T^*_{CCSI}(t)=\widehat{\cal T}_{\delta}^{\textit{bs}*}(t)-\widehat{\cal T}_{\delta}^{\textit{bf}*}(t)$
\item[8.] Repeat Steps 5 and 6 $B$ times to obtain \\ $T_{CCSI }^{*1}(t), \ldots, T_{CCSI }^{*B}(t)$.
\item[9.] Calculate the $\alpha$ and $(1-\alpha)$ quantiles,  $T_{CCSI, \alpha}^{*}(t)$ and  $T_{CCSI, (1-\alpha)}^{*}(t)$,  at each $t$. Reject $H_0$ at time $t$ if $T_{CCSI}(t)<T_{CCSI,\alpha}^{*}(t)$ or\\
 $T_{CCSI}(t)>T_{CCSI,(1-\alpha)}^{*}(t)$.
\end{enumerate}

\section*{Acknowledgements}

This work was produced as a part of the FPI grant (BES-2009-017772) from the Spanish Ministerio de Econom\'ia y Competitividad.  AM. Gonz\'alez-Montoro and R. Cao have been partially supported by the grants \\ MTM2008-00166,  MTM2011-22392 of the Spanish Ministerio de Econom\'ia y Competitividad. AM. Gonz\'alez-Montoro, N. Espinosa and J. Mari\~no have been partially supported by Xunta de Galicia (INCITE09 137 272 PR). AM. Gonz\'alez-Montoro has been  partially supported by FAPESP Research, 
Dissemination and Innovation Center for Neuromathematics (grant 2013/ 07699-0), S. Paulo Research Foundation.\\
The authors would like to thank the Action Editor and two anonymous reviewers for their interesting comments and questions. They have been of great help to improve the quality of the paper.\\



\end{document}